\documentclass[journal]{IEEEtran}

\usepackage[english]{babel}
\usepackage{color,array}
\usepackage{multirow}
\usepackage{graphicx}
\usepackage{amsmath}
\usepackage{amssymb}
\usepackage{threeparttable}
\usepackage{breqn}
\usepackage{makecell}
\usepackage{svg}
\usepackage{listings}
\usepackage{inconsolata}
\usepackage{microtype}
\usepackage{enumitem}
\usepackage{cite}

\usepackage[colorlinks,urlcolor=blue,linkcolor=blue,citecolor=blue]{hyperref}
\usepackage{tabularx}
\newcolumntype{L}[1]{>{\raggedright\arraybackslash\hsize=#1\hsize}X}
\newcolumntype{C}[1]{>{\centering\arraybackslash\hsize=#1\hsize}X}
\begin{document}

\title{From Legacy Fortran to Portable Kokkos:\\ An Autonomous Agentic AI Workflow} 

\author{
Sparsh Gupta, Kamalavasan Kamalakkannan, Maxim Moraru, Galen Shipman, and Patrick Diehl
\thanks{
Manuscript submitted \today. 
Research presented in this article was supported by the National Security Education Center (NSEC) Informational Science and Technology Institute (ISTI) using the Laboratory Directed Research and Development program of Los Alamos National Laboratory project number 20240479CR-IST. This research used resources of the National Energy Research Scientific Computing Center, a U.S. Department of Energy Office of Science User Facility operated under Contract No. DE-AC02-05CH11231. This work was also supported by the U.S. Department of Energy through the Los Alamos National Laboratory. Los Alamos National Laboratory is operated by Triad National Security, LLC, for the National Nuclear Security Administration of U.S. Department of Energy (Contract No. 89233218CNA000001). LA-UR-25-28882 
}

\thanks{Sparsh Gupta is with the Los Alamos National Laboratory, Los Alamos, NM 87544 USA and the Franklin W. Olin College of Engineering, Needham, MA 02492 USA (e-mail: sgupta1@olin.edu). Kamalavasan Kamalakkannan, Maxim Moraru, Galen Shipman, and Patrick Diehl are with the Los Alamos National Laboratory, Los Alamos, NM 87544 USA. (e-mails: \{kamalavasan,moraru,gshipman,diehlpk\}@lanl.gov)}

}

\markboth{
}
{
}

\maketitle

\begin{abstract}
Scientific applications still rely heavily on legacy Fortran codebases built for homogeneous, CPU-based systems. As High-Performance Computing (HPC) shifts toward heterogeneous GPU-accelerated architectures, many accelerators lack native Fortran bindings, creating a need to translate and optimize legacy code for portability. Frameworks like Kokkos offer performance portability through a single-source C\texttt{++} abstraction, but manual porting from Fortran to Kokkos demands significant expertise and remains time-intensive. While large language models (LLMs) show promise in source-to-source code generation, their use in autonomous workflows for translating and optimizing parallel code, particularly for performance portability across diverse hardware, is largely unexplored. This paper presents an agentic AI workflow in which specialized LLM agents collaborate to translate, validate, compile, run, test, debug, and optimize Fortran kernels into portable Kokkos C\texttt{++} programs without human intervention, bounded by configurable retry and optimization thresholds. Because such workflows are non-deterministic, we evaluate the pipeline over five runs per configuration on an NVIDIA A100, characterizing its variability in success rates, token cost, and optimization behavior. The pipeline modernizes a range of benchmark kernels, including a real-world multi-file kernel from a production climate model, into performance-portable Kokkos code across hardware partitions. Proprietary models such as GPT-5 and o4-mini-high completed the full workflow most reliably, producing optimized code that exceeded the Fortran baselines for only a few U.S. dollars, whereas open-weights models such as gpt-oss-120b and Qwen3.5-397B-A17B handled the simpler kernels but struggled with those demanding more complex Kokkos abstractions such as parallel reductions and hierarchical parallelism. This work demonstrates, at the kernel level, the feasibility of agentic AI for Fortran-to-Kokkos transformation and, more broadly, for structured software-engineering tasks such as automated code modernization and refactoring, offering a path toward autonomously modernizing legacy scientific applications across diverse supercomputers.
\end{abstract}




\section{Introduction}

\IEEEPARstart{L}{egacy} Fortran code remains the backbone of many crucial scientific computing applications across national laboratories and academic institutions, such as the WRF model~\cite{Skamarock2019-ek}, Code\_saturne~\cite{cs_ijfv_2004}, CHARMM~\cite{Brooks2009-yu}, \emph{etc}. These codes were originally designed for homogeneous, CPU-based systems and have been performance-tuned over decades for intra-node parallelism using interfaces like MPI~\cite{mpi} and OpenMP~\cite{openmp}. However, with the transition to heterogeneous High-Performance Computing (HPC) systems featuring GPU-accelerated hardware from vendors like NVIDIA, AMD, and Intel, there is a growing need to port legacy codes to run efficiently on modern architectures. Many of these accelerators lack native Fortran bindings~\cite{10.1145/3624062.3624178}, creating significant barriers to portability and performance. To address this, frameworks such as the Kokkos C\texttt{++} Performance Portability Ecosystem~\cite{KokkosEcosystem2021} have emerged. It is widely adopted for portable HPC applications, letting developers write single-source C\texttt{++} code that runs efficiently across multiple programming models like CUDA~\cite{cuda}, HIP~\cite{hip}, SYCL~\cite{sycl}, OpenMP~\cite{openmp}, HPX~\cite{Kaiser2020}, and C\texttt{++} threads. However, manually porting Fortran to Kokkos requires significant C\texttt{++} and parallel-programming expertise and is time-consuming and debugging-heavy, creating a bottleneck in modernizing legacy scientific software.

Large Language Models (LLMs) have shown strong performance on code translation, generation, and repair~\cite{chen2021evaluatinglargelanguagemodels, codex, roziere2024codellamaopenfoundation, openai2024gpt4technicalreport, poldrack2023aiassistedcodingexperimentsgpt4}. However, existing methods have not fully explored their potential for integrating diagnostics, iterative performance tuning, and hardware-specific optimization, particularly in HPC contexts where performance portability is essential. This motivates our exploration of structured, multi-agent AI workflows for modernizing legacy Fortran codebases.

In this work, we introduce an automated agentic AI workflow for translating and optimizing Fortran kernels to Kokkos C\texttt{++} across architectures. Specialized LLM agents collaborate to translate the source, validate it, manage compilation and execution on HPC clusters, fix build and runtime failures, test functionality, and propose profiler-guided optimizations. The pipeline runs without human intervention, bounded by configurable retry and optimization thresholds that determine when a stage is abandoned, and uses both proprietary LLMs via the OpenAI API and self-hosted open-weights LLMs. We evaluate it on five kernels from the NAS Parallel Benchmarks~\cite{naspb} and OpenBLAS (DGEMM)~\cite{zhang2013openblas}, plus biharmonic, a real-world multi-file kernel from the E3SM codesign suite~\cite{e3sm-model}, running each configuration five times on the NVIDIA A100 to capture the non-determinism of LLM-driven pipelines. Proprietary models GPT-5~\cite{openai_gpt5_2025} and o4-mini-high~\cite{openai_o4mini_2025} complete the full pipeline most reliably, while the open-weights gpt-oss-120b~\cite{openai2025gptoss120bgptoss20bmodel} and Qwen3.5-397B-A17B~\cite{qwen3.5_397b_a17b} handle the simpler kernels but consistently fail on those demanding more complex Kokkos abstractions such as parallel reductions and hierarchical parallelism. We show this gap manifests as specific, recurring failure modes, with gpt-oss-120b failing outright on reduction kernels and both open-weights models fragile through the iterative optimization that follows. Earlier single-run results for Llama 4 Maverick~\cite{meta2025llama4} are retained in the supplementary material. All findings in this paper are at the kernel level; complete legacy applications are outside the current scope.

\noindent Our study is guided by four research questions:
\begin{description}[leftmargin=0pt, itemindent=\parindent, labelsep=0em]
\item[\textbf{RQ1. }]\phantomsection\label{rq:portability} Can an agentic LLM workflow autonomously translate legacy Fortran kernels into functionally correct, performance-portable Kokkos C\texttt{++} across heterogeneous GPU architectures?
\item[\textbf{RQ2. }]\phantomsection\label{rq:reliability} How reliable and reproducible is the workflow across repeated runs, given the non-determinism of LLM-driven pipelines?
\item[\textbf{RQ3. }]\phantomsection\label{rq:openclosed} How do open-weights LLMs compare to proprietary models in completing the workflow, and what systematic failure modes explain the gap?
\item[\textbf{RQ4. }]\phantomsection\label{rq:optimization} How effective is profiler-guided optimization at improving the generated Kokkos code?
\end{description}

In summary, this paper makes four contributions: (i) an autonomous, profiler-in-the-loop agentic workflow that translates, compiles, runs, tests, and optimizes legacy Fortran kernels into performance-portable Kokkos C\texttt{++} across NVIDIA and AMD GPUs within configurable thresholds; (ii) a multi-run evaluation of four LLMs over six kernels, totaling 120 runs on the A100, quantifying run-to-run variability in success rates, agent invocations, token cost, and optimization trajectories; (iii) a manifest-based method for translating multi-file Fortran programs, demonstrated on a real-world E3SM kernel, that handles cross-module dependencies without program-specific build rules; and (iv) a failure-mode analysis of open-weights versus proprietary models, identifying where open-weights models break down, from outright failure on reduction kernels to fragility through iterative optimization.

The remainder of the paper covers related work (Section~\ref{sec:relatedwork}), background on Kokkos, SLURM, Spack, and Agentic AI (Section~\ref{sec:background}), the benchmark kernels (Section~\ref{sec:fortrankernels}), our methodology (Section~\ref{sec:methods}), the experimental setup (Section~\ref{sec:experimentalsetup}), results and discussion (Section~\ref{sec:results}), threats to validity (Section~\ref{sec:threats}), and conclusions and future work (Section~\ref{sec:conclusion}).


\section{Related Work}
\label{sec:relatedwork}

Researchers have explored Fortran translation for decades. Early tools like \textit{f2c}~\cite{f2c} enabled basic Fortran-to-C translation, facilitating reuse of legacy code, and source-to-source compilers such as LLNL's ROSE~\cite{quinlan2011rose} refactor Fortran HPC applications and enable semi-automatic parallelization by injecting parallel constructs (\emph{e.g.}, OpenMP or OpenACC pragmas) into legacy loops. \textit{LFortran}~\cite{lfortran}, a newer LLVM-based~\cite{llvm} open-source Fortran compiler, remains in alpha with many language features incomplete, and is not yet practical for modernizing large-scale legacy applications, let alone optimizing them.

In parallel, large language models (LLMs) have evolved from single-step code generation toward structured, agentic workflows that iteratively improve, debug, and optimize code, a direction closely tied to long-standing software goals such as automated program repair, refactoring, and validation-driven refinement~\cite{10.1145/3799693}. Multi-agent systems such as \textit{ChatDev}~\cite{qian-etal-2024-chatdev} and \textit{MetaGPT}~\cite{hong2024metagptmetaprogrammingmultiagent} coordinate role-based LLM agents through natural-language dialogue to complete full-cycle software development tasks, \textit{CodeChain}~\cite{le2024codechainmodularcodegeneration} uses a chain-of-thought methodology to incrementally generate and revise modular code with notable accuracy gains on competitive-programming benchmarks, and \cite{ashrafi2025enhancingllmcodegeneration} couples generation with execution and debugging through runtime feedback. Collectively, these works underscore the potential of structured, multi-agent automation, but they largely target general-purpose or competitive-programming software and optimize for functional correctness, rather than the performance portability and hardware-specific optimization that modernizing legacy HPC code demands.

Beyond agentic systems, recent work has assessed the standalone coding abilities of LLMs. Llama2-70B, for instance, could generate correct code and tests for simpler tasks but struggled with complex parallel workloads, often requiring manual intervention~\cite{Diehl_2025}, and other studies have probed earlier-generation models (Codex, GPT-3, Llama-2) on HPC kernels and BLAS routines under manual or one-shot prompting~\cite{e5a3cfba27bd400d83f577d889855c0a,10.1109/SCW63240.2024.00010,Godoy_2023}. The ParEval benchmark systematically evaluated LLMs on parallel code generation across OpenMP, MPI, CUDA, HIP, and Kokkos over 420 scientific problems~\cite{Nichols_2024}. Closer to our setting, several efforts translate Fortran to C\texttt{++} with LLMs: \cite{ranasinghe2025llmassistedtranslationlegacyfortran} reports compilation and functional-correctness rates across hardware platforms but finds many translations fail to compile or need manual debugging, \textit{Fortran2CPP}~\cite{chen2025fortran2cppautomatingfortrantoctranslation} uses a dual-agent dialogue to improve compile success and CodeBLEU scores, and \cite{11181523} targets multiple backends (OpenMP, OpenACC, CUDA, HIP, Kokkos) through multimodal prompting and fine-tuning with manual correction. These efforts focus on syntactic and semantic correctness under single-turn or manually guided prompting, and do not address autonomous multi-stage compilation, execution, debugging, and iterative optimization across architectures.

Despite this progress, gaps persist: static tools such as ROSE and LFortran do not optimize, while emerging LLM workflows neglect performance, portability, and feedback-driven refinement. Our work addresses both with an end-to-end agentic Fortran-to-Kokkos workflow, without fine-tuning, integrating translation, compilation, execution, functionality testing, profiling, and iterative optimization across proprietary and open-weights LLMs.


\section{Background}
\label{sec:background}

\subsection{Kokkos}

Kokkos~\cite{KokkosEcosystem2021} lets developers write single-source C\texttt{++} code by abstracting both the execution space (\emph{e.g.}, GPU threads vs. CPU cores) and the memory space (\emph{e.g.}, device vs. host memory), so the same logic compiles and runs across diverse architectures. It provides high-level parallel patterns, including \lstinline{parallel_for} for parallel loops, \lstinline{parallel_reduce} for reductions, and \lstinline{parallel_scan} for prefix computations, and supports \textit{hierarchical parallelism} through thread teams (which synchronize and share scratch-pad memory) and vector lanes. It also exposes memory-layout policies for common data structures, making it a comparatively complete ecosystem for performance portability across GPU architectures.

\subsection{SLURM}

SLURM~\cite{osti_15002962} (Simple Linux Utility for Resource Management) is an open-source workload manager widely used to schedule jobs and allocate resources on HPC clusters. Our pipeline uses SLURM to submit and monitor the build and run jobs, but its modular design allows easy adaptation to other schedulers if needed.

\subsection{Spack}

Spack~\cite{spack_sc15} is an open-source HPC package manager that enables reproducible, portable software environments across architectures. In our workflow, it manages the compilers and backend-specific Kokkos builds, which is essential for reproducibility and for targeting different GPU backends from a single pipeline.

\subsection{Agentic AI}

Agentic AI systems coordinate multiple LLM agents that operate autonomously to accomplish complex tasks through structured decision-making, tool use, and inter-agent communication. Unlike one-shot prompting, each agent maintains state, calls external tools, and produces structured output, and can be specialized to a role so that a complex workflow is decomposed into subtasks delegated across agents. This mirrors established software-engineering principles: each agent carries a single responsibility with clear separation of concerns, while bounded-retry error handling and versioned intermediate artifacts give the pipeline the robustness and reproducibility of a well-engineered system.

Several frameworks, such as LangChain, CrewAI, and Microsoft AutoGen~\cite{wu2024autogen}, provide abstractions for tool integration, planning, and agent coordination. We build on the open-source OpenAI Agents SDK~\cite{openai_agents_sdk}, a lightweight Python framework that wraps functions as tools (letting agents drive compilers, file systems, job schedulers, and profilers), supports per-agent prompts and memory, and enables inter-agent communication. Crucially, it routes requests across providers, so agents can invoke OpenAI-hosted or self-hosted models without changing the workflow logic.


\section{Benchmark Fortran Kernels for Evaluation}
\label{sec:fortrankernels}

In this work, we evaluate our pipeline on six Fortran kernels, chosen for their parallel complexity, diverse functionalities, and, in one case, real-world multi-file structure. To ensure consistent benchmarking with the generated Kokkos code, we modularized each kernel as a standalone Fortran subroutine with pre-initialized values that accepts two input parameters (same as the Kokkos programs): the problem size ($n$) and the number of kernel repetitions. These subroutines were wrapped in minimal driver programs for execution, testing, and runtime benchmarking, and only the subroutines (without the driver) are passed to the workflow for translation. Five kernels (CG, EP, MG, FT, and DGEMM) are self-contained single-file programs from standard HPC benchmark suites, while biharmonic is a multi-file kernel from a production application, used to test the cross-file dependencies typical of real legacy codebases. During development we also validated the pipeline on the STREAM kernels~\cite{McCalpin1995, McCalpin2007}, but they are simple and not used in the quantitative evaluation. A summary is provided in Table~\ref{tab:benchmarks}.

\begin{table}[htbp]
\centering
\scriptsize
\begin{threeparttable}
\caption{Benchmarks used in this work}
\label{tab:benchmarks}
\begin{tabular}{|l|c|c|}
\hline
\textbf{Kernel} & \textbf{Type} & \textbf{Modularized Subroutine Lines of Code} \\
\hline
CG             & Memory-bound  & 165 \\
\hline
EP             & Compute-bound & 129 \\
\hline
MG             & Memory-bound  & 139 \\
\hline
FT             & Memory-bound  & 230 \\
\hline
DGEMM          & Compute-bound & 177 \\
\hline
BIHARMONIC     & Memory-bound  & 256 (132 + 124)\tnote{*} \\
\hline
\end{tabular}
\begin{tablenotes}
\item[*] \scriptsize Multi-file kernel: main routine (\texttt{biharmonic.f}) plus the \texttt{kernel_utils.f} module.
\end{tablenotes}
\end{threeparttable}
\end{table}

\subsection{NAS Parallel Benchmarks (NPB)}

The NPB3.4~\cite{naspb} are a suite of performance kernels developed by the NASA Advanced Supercomputing (NAS) Division to evaluate the performance of highly parallel supercomputers.

\subsubsection{Conjugate Gradient (CG)}
The CG benchmark estimates the smallest eigenvalue of a large sparse symmetric positive-definite matrix. It uses inverse iteration, applying the conjugate gradient method to solve linear systems $Ax = b$,
where $A$ is a sparse symmetric positive-definite matrix, $x$ is the unknown vector, and $b$ is a known vector, exercising unstructured memory access, indirect indexing, and irregular loop structures.

\subsubsection{Embarrassingly Parallel (EP)}
The EP kernel generates independent Gaussian random variates using the Marsaglia polar method and estimates values such as the integral of the Gaussian probability distribution.
\begin{equation}
    x_1 = \sqrt{\frac{-2 \ln(u_1)}{u_1^2 + u_2^2}} u_1, \hspace{3mm} x_2 = \sqrt{\frac{-2 \ln(u_1)}{u_1^2 + u_2^2}} u_2,
\label{eq:ep}
\end{equation}

where $u_1, u_2$ are uniformly distributed in $(-1, 1)$ and $x_1, x_2$ are the resulting Gaussian variates. It involves no inter-task communication, representing purely parallel, compute-intensive workloads.

\subsubsection{Multi-Grid (MG)}
The MG kernel approximates the solution to a three-dimensional discrete Poisson equation using a V-cycle multigrid method, $\nabla^2 u(x, y, z) = v(x, y, z)$, where $\nabla^2$ denotes the 3D Laplacian operator and $u$, $v$ are the unknown solution and source terms. The method approximates the solution across multiple resolution levels, with regular memory access and hierarchical computation typical of PDE solvers.

\subsubsection{Fourier Transform (FT)}
The FT kernel solves a three-dimensional partial differential equation (PDE) using the Fast Fourier Transform (FFT), computed as:
\begin{dmath}
    F(k_x, k_y, k_z) = \sum^{N_x - 1}_{x=0} \sum^{N_y - 1}_{y=0} \sum^{N_z - 1}_{z=0} f(x,y,z) e^{-2 \pi i (\frac{k_x x}{N_x} + \frac{k_y y}{N_y} + \frac{k_z z}{N_z})},
\end{dmath}
where $F(k_x, k_y, k_z)$ are the frequency-domain coefficients of the spatial data $f(x,y,z)$, and $N_x, N_y, N_z$ are the grid sizes. It combines structured memory access with global communication, typical of PDE solvers and signal-processing workloads.

\subsection{DGEMM from OpenBLAS}

DGEMM (Double-precision GEneral Matrix Multiplication) is a fundamental OpenBLAS~\cite{zhang2013openblas} routine for double-precision matrix-matrix multiplication $C = \alpha A B + \beta C$, where $A, B, C$ are matrices, and $\alpha, \beta$ are scalar multipliers. This kernel is ideal for evaluation as it is both compute-bound and cache-intensive, with performance strongly tied to memory layout, loop ordering, and tiling.

\subsection{Biharmonic Operator (E3SM)}

To evaluate the pipeline on a kernel drawn from a production scientific application, we include the weak-form biharmonic kernel from the codesign-kernels suite~\cite{E3SM_CODESIGN} of the Energy Exascale Earth System Model (E3SM)~\cite{e3sm-model, E3SM}, extracted from its atmospheric dynamical core. The biharmonic operator applies the Laplacian twice and is used for hyperviscosity-based numerical diffusion in spectral-element atmospheric dynamics, $\nabla^4 u = \nabla^2\!\left(\nabla^2 u\right)$, where $\nabla^2$ is the Laplacian operator and $u$ is the field to which scale-selective diffusion is applied. Operating over a set of spectral elements, each with $\mathrm{NP}\times\mathrm{NP}$ nodes across multiple vertical levels and tracers, the kernel combines a differentiation matrix with element metric, mass, and tensor-viscosity terms to compute the diffusion update. Unlike the NPB and DGEMM kernels, biharmonic is a \emph{multi-file} program: its main routine (\texttt{biharmonic.f}) depends on a helper module (\texttt{kernel\_utils.f}) through Fortran \texttt{use}-association, exercising the cross-file dependencies characteristic of real scientific codebases. To handle such multi-file inputs, we extend the pipeline with a manifest-based packaging scheme, described in Section~\ref{sec:methods}.


\section{Methods}
\label{sec:methods}

\begin{figure*}[htbp]
    \centering
    \includegraphics[width=\linewidth,trim= 0 475 0 285,clip]{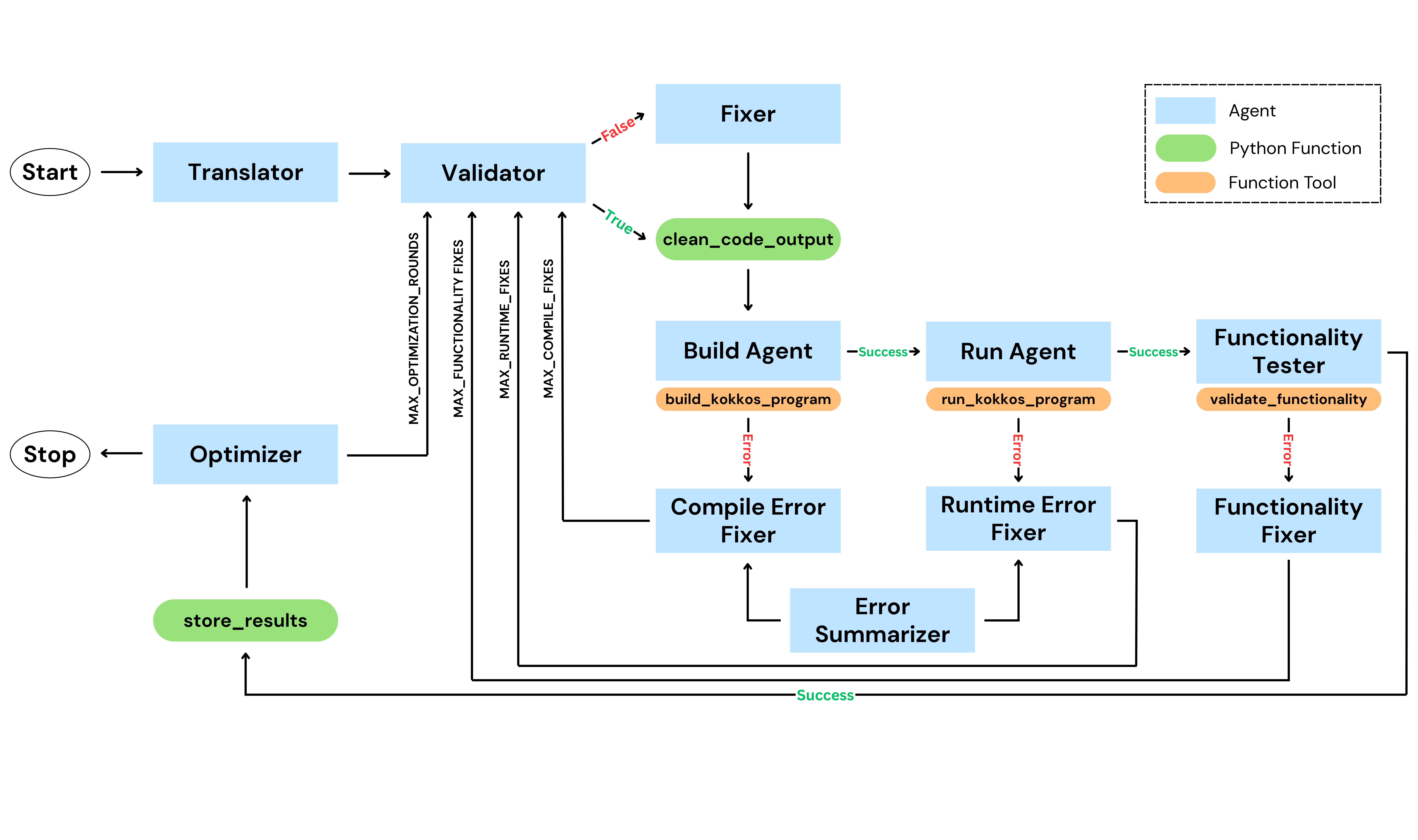}
    \caption{Agentic AI workflow for autonomous Fortran-to-Kokkos translation, validation, compilation, runtime execution, functionality testing, and performance optimization. Fixer Agents are triggered on error events (\emph{e.g.}, failed compilation, runtime fault, or incorrect functionality testing output). Agent invocation limits at each stage are enforced via configurable thresholds (\emph{e.g.}, \texttt{MAX\_COMPILE\_FIXES}). Function tools invoked by the Build and Run agents (\lstinline{build\_kokkos\_program}, \lstinline{run\_kokkos\_program}) utilize SLURM to schedule and monitor jobs and Spack to load the correct Kokkos environment on the hardware partitions. All artifacts and metrics are versioned and stored per version run using the \lstinline{store\_results} function.}
    \label{fig:workflow}
\end{figure*}

The proposed workflow (Fig.~\ref{fig:workflow}) is an autonomous pipeline of specialized LLM agents that translates, builds, runs, tests, and optimizes legacy Fortran kernels into performance-portable Kokkos C\texttt{++}, handling build failures, runtime errors, and correctness mismatches through iterative agent invocations.

\subsection{Agents and Roles}

The workflow is built from modular, role-specific agents that operate sequentially within each stage, collaborating via intermediate outputs. We group them into four phases, as outlined below.

\subsubsection{Translation and Validation}

The pipeline begins with a \textit{Translator Agent} that converts legacy Fortran kernels into standalone Kokkos C\texttt{++} programs, preserving the exact computational semantics, variable names, and function signatures while ensuring portability across Kokkos backends such as CUDA, HIP, and OpenMP. Several directives were added to the prompt after observing recurring failures. The translator must call \lstinline{Kokkos::fence()} exactly once inside \lstinline{main()} after all parallel work and before measuring or printing results, since the agent would otherwise omit it; this also gives a consistent point at which functionality-testing code is later injected. It is directed to implement all computational logic fully, as outputs would otherwise leave placeholder comments for the main computation. It must declare all \lstinline{Kokkos::View} objects in an inner scope of \lstinline{main()} that closes before \lstinline{Kokkos::finalize()}, since Views destroyed after finalization trigger a runtime error. Finally, device results needed on the host are transferred through mirror views via \lstinline{Kokkos::create_mirror_view_and_copy} (or an explicit \lstinline{deep_copy}) rather than by dereferencing device Views directly, which would fail on non-host backends. The translated program supports command-line arguments for input size ($n$) and kernel repetitions, runs the computational loop the requested number of repetitions, and prints the total execution time to six decimal places.

The generated code is then evaluated by a \textit{Validator Agent}, which verifies that the output is syntactically valid C\texttt{++} and free from any non-code content such as comments, markdown, or natural language. If the code fails validation, it is passed to a \textit{Fixer Agent}, which applies structural and syntactic corrections if needed and removes any non-code texts while preserving the computational logic. The output is usually wrapped in code-block tags (\lstinline{```cpp} and/or \lstinline{```}), so a python function (\lstinline{clean_code_output}) extracts the compilation-ready C\texttt{++} from within them via a regular expression.

\subsubsection{Compilation and Execution}

Once the code is obtained from the last phase, it is compiled via the \textit{Build Agent}, which invokes a wrapped python-function tool (\lstinline{build_kokkos_program}) to submit a SLURM job that compiles the program using architecture-specific toolchains on the desired hardware partition. This SLURM job utilizes Spack to load the correct software environment, including the appropriate version of Kokkos and backend libraries/packages. The output of the build process is monitored, and in case of failure, the \textit{Error Summarizer Agent} parses the stderr logs to extract the root cause and condense it into a concise plain text summary not exceeding 20 lines. If possible, it also suggests a couple potential fixes for the issues. This summary is then utilized by the \textit{Compile Error Fixer}, which attempts to correct the C\texttt{++} code based on the diagnosed issues.

Following successful compilation, the code is executed using the \textit{Run Agent}, which invokes a wrapped python-function tool (\lstinline{run_kokkos_program}) with the desired configuration and parameters and submits a runtime SLURM job on the desired hardware partition. The same Spack-managed environment is reused during execution to guarantee that runtime libraries, compiler toolchains, and backend configurations are consistent with those used at build time. GPU profiling is also enabled and is run for the maximum input size. Similar to the build phase, any runtime failures are summarized by the \textit{Error Summarizer} and addressed by the \textit{Runtime Error Fixer}, which patches the code based on runtime-specific errors such as segmentation faults, invalid memory access, or device synchronization issues. 

\subsubsection{Functionality Testing}

After successful execution, the functional correctness of the translated program is evaluated by the \textit{Functionality Tester} which utilizes another wrapped python-function tool (\lstinline{validate_functionality}). This agent injects kernel-customized testing code into the Kokkos program, immediately after \lstinline{Kokkos::fence()}, to capture the output of the relevant resultant array of the program into a CSV file. Then, this code is re-compiled to ensure it is executable after the code injection. Simultaneously, the Fortran kernel's driver program is also compiled using GFortran with the required flags for OpenMP, etc. Once the builds are successful, the agent runs both the original Fortran kernel program and the translated Kokkos version (with GPU profiling disabled) on the same inputs and compares their outputs across a small sweep of problem sizes. For all kernels except the EP kernel, we check if the outputs mismatch by exceeding a tolerance limit. For the EP kernel, due to the pseudo-randomness associated with the results, we just check for a non-zero output instead. For biharmonic, the original order-dependent random initialization gave the Fortran and Kokkos versions different inputs (since they fill arrays in different orders), so we seed each array deterministically by element index so both sides operate on identical inputs before the tolerance check. In case of functionality incorrectness, the generated Kokkos code and the original Fortran code is passed to the \textit{Functionality Fixer}, which attempts to correct logic-level issues while preserving performance-portable constructs and overall structure. Each corrected version undergoes re-compilation and re-execution before re-testing. The injected code is removed from the Kokkos program once this process is finished.

\subsubsection{Optimization}

\begin{figure}[htbp]
    \centering
    \includegraphics[width=\linewidth,trim= 0 1530 0 0,clip]{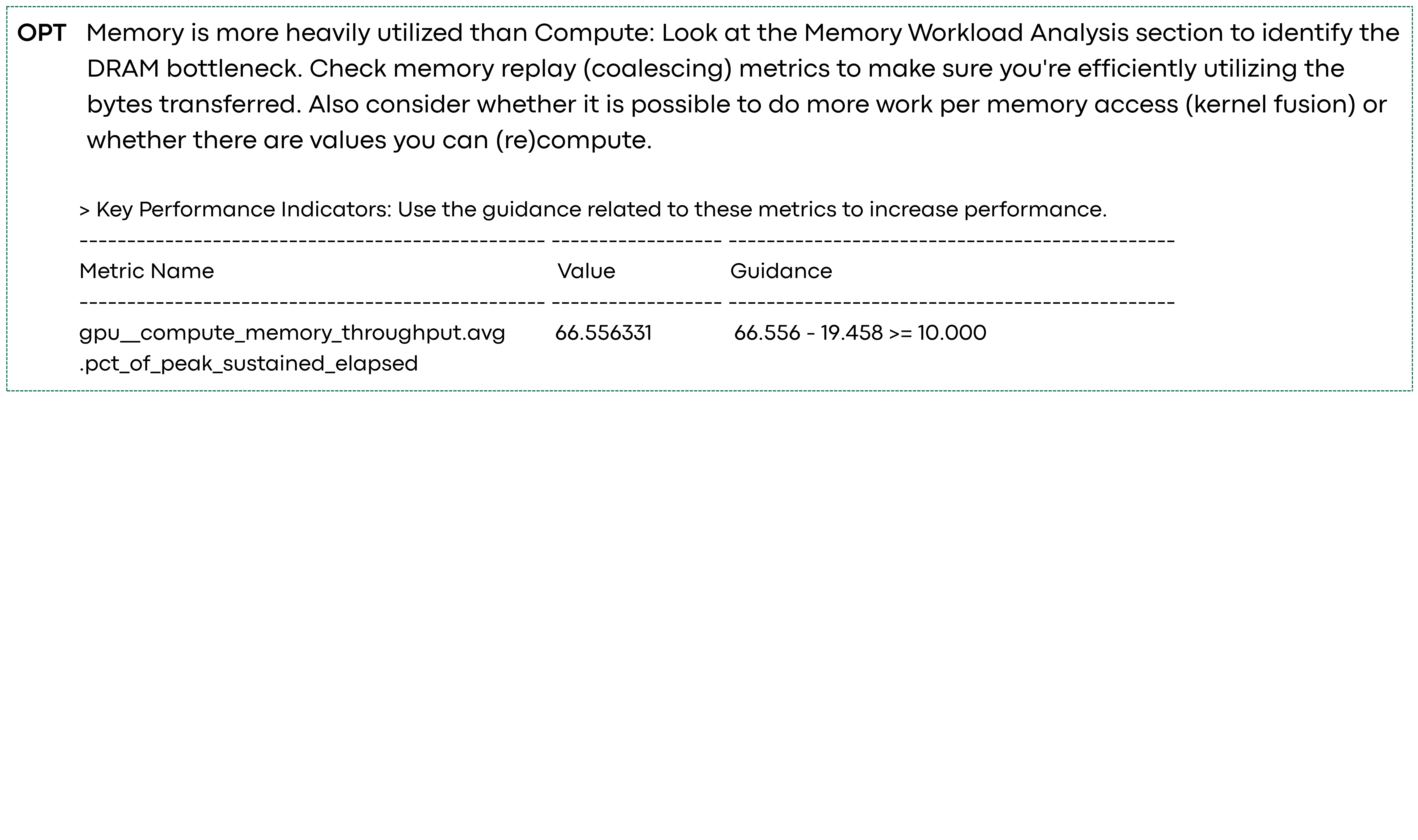}
    \caption{Example NVIDIA Nsight Compute OPT suggestion}
    \label{fig:ncu-opt}
\end{figure}

Once a functionally correct baseline is established, the pipeline enters the optimization loop. The \textit{Optimizer Agent} is provided with profiling summaries generated via hardware GPU profilers (\emph{e.g.}, NVIDIA Nsight Compute or AMD ROCProfiler). For the profiling report generated by NVIDIA Nsight Compute (NCU), we parse out the suggested OPT points (\emph{e.g.}, Fig. \ref{fig:ncu-opt}) which highlight potential optimizations and estimated speedup for the relevant kernel loop and combine it into a summary. AMD rocprofv3 produces no such report, so we summarize metrics like memory-access statistics, cache hit/miss rates, and occupancy into a standard diagnostic summary based on thresholds established for improving these metrics. Based on this feedback, the agent applies structural code changes aimed at improving memory layout, loop ordering, execution policies (\emph{e.g.}, block sizes, vector lengths), and use of Kokkos hierarchical parallelism constructs such as \lstinline{TeamPolicy} and \lstinline{ThreadVectorRange}. Each optimized version is recompiled, re-executed, re-tested, and re-profiled to determine runtime and performance improvements. All functionally-correct versions produced during the optimization rounds are retained, not just the final one. Because profiler-guided edits do not always improve performance, the fastest functionally-correct version in a given run is not necessarily the last; retaining every version lets us identify the best-performing one per run and report the resulting runtime distribution across runs in Section~\ref{sec:results}. The Build, Run, Functionality Tester, and Fixer agents are reused here for consistency across optimization iterations. 

\subsection{Multi-File Kernel Handling}
\label{sec:multifile}

The five single-file kernels are passed to the pipeline as a single Fortran source. Real scientific codebases, however, are often split across multiple files connected by Fortran \texttt{use}-association, as in the biharmonic kernel (Section~\ref{sec:fortrankernels}). To support such multi-file inputs without modifying the agent prompts, we add a lightweight manifest-based packaging scheme.

Each kernel directory may contain a \texttt{manifest.json} file that declares the source files in dependency order together with the entry routine. For the biharmonic kernel, this lists \texttt{kernel\_utils.f} before \texttt{biharmonic.f}, with \texttt{biharmonic\_kernel} as the entry routine. When a manifest is present, the Fortran benchmarking tool (\lstinline{fortran_benchmark.py}) reads the file list and compiles the sources with GFortran in the declared order, ensuring that modules are built before the units that depend on them. When no manifest is present, the tool falls back to compiling a single \texttt{<kernel>.f}, so the single-file kernels are unaffected. 

For translation, the multi-file source is presented to the Translator Agent as one concatenated input, with each file preceded by a \lstinline{! FILE: <name>} delimiter comment in dependency order. The translator is instructed to emit a single self-contained Kokkos C\texttt{++} program that preserves the cross-module semantics of the original, inlining the helper module's functionality rather than reproducing the file structure. This keeps the downstream compilation, execution, functionality-testing, and optimization stages identical to the single-file case.
Because the mechanism is driven entirely by the manifest, extending the pipeline to a new multi-file kernel requires only adding its files and a manifest to the kernel directory, with no changes to the driver logic or agent prompts.

\subsection{Workflow Orchestration}

The agentic pipeline coordinates agents sequentially across stages: translation, validation, compilation, execution, functionality testing, and optimization. Each stage is an asynchronous, isolated function that invokes one or more agents and tools, so any stage can be extended, replaced, or reused independently. The orchestration enforces user-defined fix-attempt thresholds for robustness: a stage aborts after reaching \texttt{MAX\_COMPILE\_FIXES}, \texttt{MAX\_RUNTIME\_FIXES}, or \texttt{MAX\_FUNCTIONALITY\_FIXES}, and optimization continues until \texttt{MAX\_OPTIMIZATION\_ROUNDS}. All artifacts and metrics, including versioned code, runtimes, fix counts, and token usage, are logged per run, with agent traces recorded via MLflow~\cite{mlflow} and cleanup routines maintaining isolation across experiments. This combination of asynchronous execution, version tracking, and bounded iteration lets the workflow run autonomously within its configured limits and abort cleanly once they are reached.


\section{Experimental Setup}
\label{sec:experimentalsetup}

\subsection{Hardware and System Configuration}

\begin{table}[htbp]
\centering
\scriptsize
\setlength{\tabcolsep}{4pt}
\renewcommand{\arraystretch}{1.2}
\caption{Hardware Specifications}
\label{tab:hardware}
\begin{tabular}{|c|c|c|c|}
\hline
\textbf{GPU} & \textbf{CPU} & \textbf{Arch.} & \makecell{\textbf{CPU} \\ \textbf{Cores}} \\
\hline
NVIDIA A100, 40\,GiB HBM2 & AMD EPYC 7763 & x86\_64 & 64 \\
\hline
AMD MI250, 64\,GB HBM2e & $2\times$ EPYC 7763 & x86\_64 & 128 \\
\hline
NVIDIA GH200, 96\,GB HBM3 & ARM Neoverse-V2 & aarch64 & 72 \\
\hline
\end{tabular}
\end{table}

\begin{table}[htbp]
\centering
\caption{Software Versions}
\begin{tabular}{|l|c|l|c|}
\hline
\textbf{Software} & \textbf{Version} & \textbf{Software} & \textbf{Version} \\
\hline
Spack   & 1.0.0.dev0   & GCC     & 8.5.0 \\
\hline
CMake   & 3.31.6       & Kokkos  & 4.6.01 \\
\hline
CUDA    & 12.6.3       & HIP     & 6.4.0 \\
\hline
Python  & 3.11.13      & OpenAI Agents SDK & 0.1.0 \\
\hline
LiteLLM & 1.73.6       & Ollama  & 0.11.4 \\
\hline
\end{tabular}
\label{tab:software}
\end{table}

All experiments ran on GPU-accelerated HPC nodes with AMD or NVIDIA architectures (Table~\ref{tab:hardware}). The A100 partition served as the primary platform for the full multi-run evaluation matrix reported in this paper, while the MI250 and GH200 partitions were used for single-run experiments demonstrating cross-architecture portability of the generated Kokkos code, with detailed per-architecture results provided in the supplementary material. Kernels were compiled and executed using SLURM-managed jobs, and architecture-specific toolchains were selected using Spack-based environments. Software versions utilized in the workflow are listed in Table~\ref{tab:software}.

\subsection{LLM Inference Setup}

\begin{table}[htbp]
\centering
\scriptsize
\begin{threeparttable}
\caption{LLMs Used}
\label{tab:llms}
\begin{tabular}{|l|c|c|c|c|}
\hline
\textbf{Model} & \textbf{Provider} & \textbf{Type} & \makecell{\textbf{Size} \\ \tiny{(parameters)}} & \makecell{\textbf{Cost} \\ \tiny{(in/out, per 1M tokens)}} \\
\hline
GPT-5~\cite{openai_gpt5_2025} & OpenAI & Proprietary & Unknown & \$1.25 / \$10 \\
\hline
o4-mini-high~\cite{openai_o4mini_2025} & OpenAI & Proprietary & Unknown & \$1.10 / \$4.40 \\
\hline
\makecell[l]{gpt-oss-120b\\~\cite{openai2025gptoss120bgptoss20bmodel}} & OpenAI & \makecell{Open-\\weights} & \makecell{117B total \\ 5.1B active} & - \\
\hline
\makecell[l]{Qwen3.5-397B\\-A17B~\cite{qwen3.5_397b_a17b}} & Alibaba & \makecell{Open-\\weights} & \makecell{397B total \\ 17B active} & - \\
\hline
\makecell[l]{Llama 4\\Maverick~\cite{meta2025llama4}}\tnote{$\dagger$} & Meta & \makecell{Open-\\weights} & \makecell{400B total \\ 17B active} & - \\
\hline
\end{tabular}
\begin{tablenotes}
\item[$\dagger$] \scriptsize Used in the supplementary single-run cross-architecture experiments.
\end{tablenotes}
\end{threeparttable}
\end{table}

To benchmark different LLMs with our workflow, we employed both proprietary and open-weights models as listed in Table~\ref{tab:llms}, without any LLM fine-tuning. We use the models as-is because fine-tuning is outside the scope of this feasibility study. The main multi-run evaluation matrix used GPT-5, o4-mini-high, gpt-oss-120b, and Qwen3.5-397B-A17B, while Llama 4 Maverick was used in our earlier single-run cross-architecture experiments, whose results are provided in the supplementary material. The proprietary OpenAI models were accessed via the official OpenAI API and incur usage-based token costs, which we report in our results. The open-weights models were self-hosted on the HPC nodes via Ollama~\cite{ollama}, a containerized inference engine, and therefore incur no per-token usage cost; requests were routed through LiteLLM~\cite{litellm}, a unified inference proxy providing a consistent API interface that enables transparent switching between model endpoints without changing the underlying workflow logic. Qwen3.5-397B-A17B was run with its thinking mode disabled to match the lightweight default reasoning footprint of gpt-oss-120b.

\subsection{Runtime Configuration}

\begin{table}[htbp]
\centering
\scriptsize
\caption{Runtime Settings for the A100 Evaluation Matrix}
\begin{tabular}{|l|c|c|c|c|c|}
\hline
\textbf{Kernel}
& \makecell{\textbf{MIN\_N}}
& \makecell{\textbf{MAX\_N}}
& \makecell{\textbf{Num. of} \\ \textbf{Sizes ($n$)}}
& \makecell{\textbf{Program} \\ \textbf{Iters.}}
& \makecell{\textbf{Kernel} \\ \textbf{Reps ($r$)}} \\
\hline
CG & 1000 & 1000000 & 10 & 2-5 & 500 \\
\hline
EP & 18 & 28 & 5 & 2-5 & 50 \\
\hline
MG & 32 & 256 & 10 & 2-10 & 250 \\
\hline
FT & 32 & 128 & 5 & 2-5 & 50 \\
\hline
DGEMM & 1024 & 4096 & 5 & 2-5 & 5 \\
\hline
BIHARMONIC & 64 & 2048 & 5 & 2-5 & 100 \\
\hline
\end{tabular}
\label{tab:runtime}
\end{table}

Each translated kernel was executed across a number of input sizes ($n$), sampled uniformly or logarithmically between a configurable minimum and maximum input size, \texttt{MIN\_N} and \texttt{MAX\_N} respectively, depending on the kernel. For each problem size, the kernel computation was executed for multiple repetitions ($r$) within each run to capture stable timing measurements. For some kernels the number of iterations was scaled inversely with input size (\emph{i.e.}, fewer iterations for larger n) to keep total job runtimes within a 30-minute wall-clock limit, while for others a fixed number of iterations was used across all input sizes. This strategy normalized total compute time across large and small input sizes while preserving computational efficiency. These runtime parameters were kept consistent across LLM variants to enable fair comparison. GPU profiling was enabled at \texttt{MAX\_N} in each run to extract performance diagnostics. Each kernel was run for 5 \texttt{MAX\_OPTIMIZATION\_ROUNDS} after the baseline run. Fix attempts for each version run during the agentic workflow were bounded by configurable thresholds: up to 10 \texttt{MAX\_COMPILE\_FIXES}, 10 \texttt{MAX\_RUNTIME\_FIXES}, and 10 \texttt{MAX\_FUNCTIONALITY\_FIXES}. Table~\ref{tab:runtime} lists the runtime parameters for the multi-run A100 evaluation matrix; the settings for the supplementary single-run experiments on the MI250 and GH200 partitions are provided in the supplementary material.

\subsection{Optimization Evaluation}

To compare optimization progress across versions, we report GFLOPS at the maximum input size for the five kernels with closed-form FLOP models, and wall-clock speedup for biharmonic (explained below). GFLOPS is computed as:
\begin{equation}
    GFLOPS = \frac{FLOPS(n,\hat{i},r)}{t_{kernel} \times 10^9},
\end{equation}
where $n$ is the input size, $\hat{i}$ is the effective per-size iteration count after scaling, $r$ is the kernel repetitions per run, and $t_{kernel}$ is the wall-clock runtime over the full loop. FLOPS use closed-form models calibrated to NPB-3 class-A conventions where applicable:
\begin{itemize}
    \item CG: sparse SpMV on the \texttt{makea()} matrix ($nnz = 43n$, inner-CG cap $c_{max}=25$); $\mathrm{FLOPS} = r(2\,nnz + 3n + c_{max}(2\,nnz + 10n)) + (2\,nnz + 3n)$.
    \item EP: polar method (RNG plus $\pi/4$ accept rate); $\mathrm{FLOPS} = r(50\cdot 2^{n+1} + 8(\tfrac{\pi}{4})2^n) \approx 106\,r\cdot 2^n$, excluding transcendentals.
    \item FT: $\mathrm{FLOPS} = r(5n^3 \log_2 n^3)$.
    \item MG: V-cycle stencil work; $\mathrm{FLOPS} = r(576\,n^3)$.
    \item DGEMM ($\alpha=1$, $\beta=2$); $\mathrm{FLOPS} = r(2n^3 + 3n^2)$.
\end{itemize}
For CG, EP, MG, and FT, $\hat{i}$ scales inversely with $n$ down to a floor of 2; DGEMM and biharmonic use the configured maximum at $n_{min}$ and 2 thereafter. This GFLOPS is an \emph{algorithmic} work-rate, invariant across translations of the same algorithm and thus suited to tracking optimization gains; hardware-measured rates (Table~\ref{tab:roofline_analysis}) can exceed it by $2$--$3\times$ from SASS-level transcendental expansion and FMA counting.

For biharmonic we instead report wall-clock speedup over the first version, $S_v = t_{v_1}/t_v$ at the maximum input size. GFLOPS is unreliable here: the measured time is dominated by per-repetition allocation and initialization rather than compute, and the HOMME-style element-local operators have no canonical NPB FLOP count, leaving the closed-form estimate and hardware-measured rate roughly $1000\times$ apart. Wall-clock speedup sidesteps both and directly answers whether a version beats the baseline.


\section{Results and Discussion}
\label{sec:results}

\begin{figure}[htbp]
    \centering
    \includegraphics[width=\linewidth,trim= 0 0 0 80,clip]{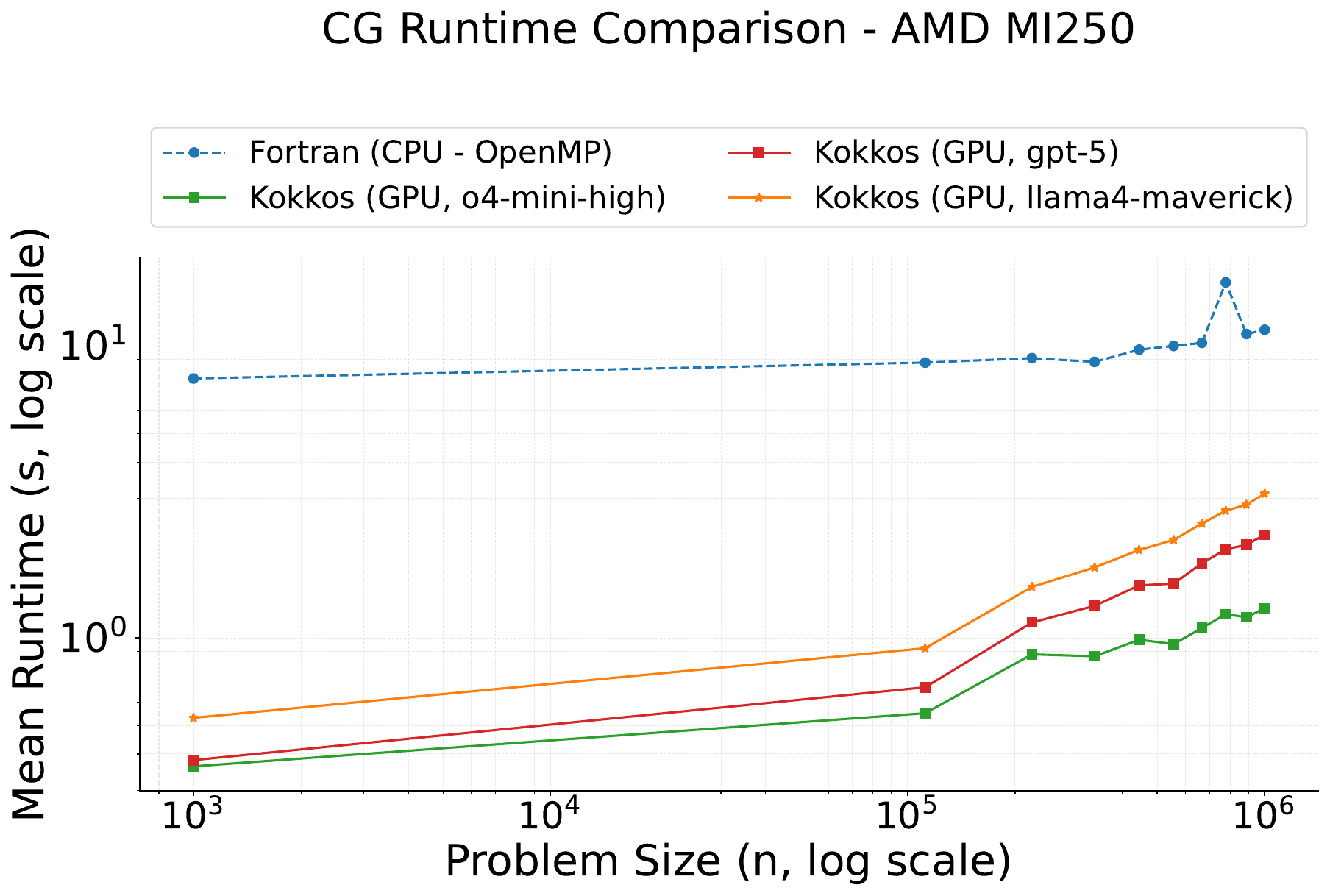}
    \caption{Mean runtime of the CG kernel on the AMD MI250 against problem size on log-log axes, comparing the Fortran CPU (OpenMP) baseline with the best Kokkos GPU version from each model in the supplementary single-run cross-architecture experiments: GPT-5, o4-mini-high, and Llama 4 Maverick.}
    \label{fig:runtime_comparison_cg_mi250}
\end{figure}

Unless otherwise noted, all results in this section come from the multi-run evaluation on the NVIDIA A100, in which each of the four LLMs (GPT-5, o4-mini-high, gpt-oss-120b, Qwen3.5-397B-A17B) was run five times on each of the six kernels, for a total of 120 runs. Reporting five runs per configuration lets us characterize the run-to-run variability inherent to LLM-driven pipelines rather than drawing conclusions from a single execution. The figures report means or medians across the five runs together with their spread, and the number of runs that reached each stage is shown wherever runs were lost to threshold exhaustion. The MI250 and GH200 partitions were evaluated in a separate set of single-run experiments that demonstrate cross-architecture portability of the generated code; one of the AMD MI250 results is shown in Fig.~\ref{fig:runtime_comparison_cg_mi250}, and the full per-architecture results, including the Llama 4 Maverick model, are provided in the supplementary material. 

\begin{figure*}[htbp]
\centering
    \includegraphics[width=\linewidth,trim= 0 0 0 10,clip]{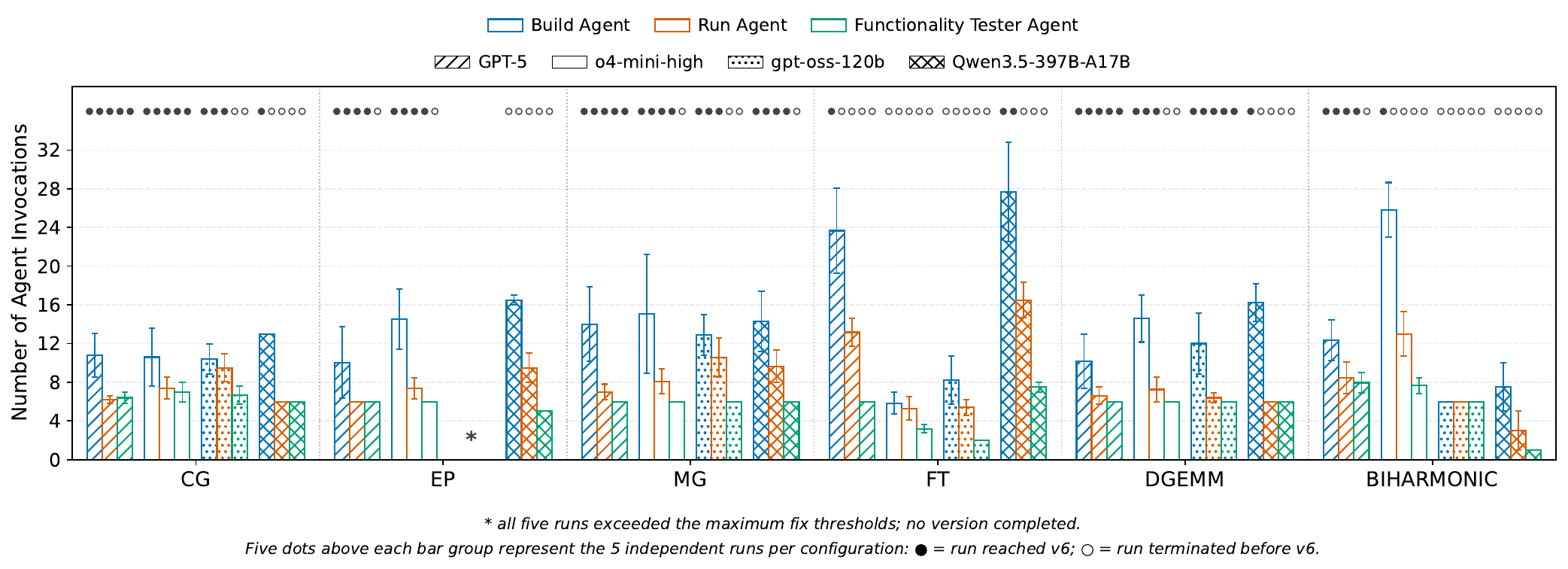}
    \caption{Number of Build, Run, and Functionality Tester agent invocations per kernel on the NVIDIA A100, grouped by LLM within each kernel. Each version's contribution to a bar is averaged over the runs that reached that version, so runs that terminate early contribute to fewer versions; error bars denote $\pm$1 standard deviation. The five dots above each group show how many of the five runs completed every optimization round through v6 (filled) or terminated earlier (open). An asterisk marks EP with gpt-oss-120b, where no run produced a functionally correct baseline, so no bars are shown.}
    \label{fig:agent_invocations}
\end{figure*}

Fig.~\ref{fig:agent_invocations} summarizes the agent invocations required to complete the workflow on the NVIDIA A100, grouped by LLM within each kernel. The proprietary models produce a functionally correct baseline and optimize it on every kernel, with the build agent invoked most often since compilation errors are the most common failure the pipeline has to repair, while the functionality tester count stays low and stable, close to one per version, because functional correctness was usually confirmed in a single attempt once a version compiled and ran. The dots show that reaching the final version is itself kernel-dependent: on FT even GPT-5 and o4-mini-high rarely complete all five optimization rounds, since later attempts tend to exhaust the fix thresholds, though the pipeline retains the fastest correct version from each run, which as the trajectories show is frequently an earlier one rather than v6.

The open-weights models show both higher invocation counts and far more early termination, visible as a preponderance of open circles on the harder kernels. Qwen3.5-397B-A17B, when it proceeds, often requires the largest number of build-agent invocations of any model, with wide error bars on kernels such as CG and FT, indicating that it repeatedly triggers and repairs compilation failures before converging. gpt-oss-120b fails EP outright, with all five runs exceeding the fix thresholds without producing even a baseline, marked by the asterisk and the absence of bars. EP requires a parallel reduction, and this total failure on a reduction kernel is the first of the failure modes that recur across the per-kernel results: both open-weights models break on specific Kokkos abstractions. A low bar with mostly open circles therefore reflects truncated workflows rather than efficient completion, so bar heights should be read together with the dot indicators.

\begin{table}[htbp]
\centering
\scriptsize
\begin{threeparttable}
\caption{Per-model success rates across the 120-run A100 evaluation matrix (six kernels, five runs each, 30 runs per model).}
\label{tab:success_rates}
\begin{tabular}{|l|c|c|c|}
\hline
\textbf{Model} & \textbf{Type} &
\makecell{\textbf{Functional} \\ \textbf{baseline}\tnote{*}} &
\makecell{\textbf{Full workflow} \\ \textbf{completed}\tnote{$\dagger$}} \\
\hline
GPT-5             & Proprietary  & 29/30 (97\%) & 24/30 (80\%) \\
\hline
o4-mini-high      & Proprietary  & 28/30 (93\%) & 17/30 (57\%) \\
\hline
gpt-oss-120b      & Open-weights & 19/30 (63\%) & 11/30 (37\%) \\
\hline
Qwen3.5-397B-A17B & Open-weights & 19/30 (63\%) &  8/30 (27\%) \\
\hline
\end{tabular}
\begin{tablenotes}
\item[*] \scriptsize Runs that produced at least a functionally correct v1 baseline.
\item[$\dagger$] \scriptsize Runs that completed all optimization rounds through v6.
\end{tablenotes}
\end{threeparttable}
\end{table}

Table~\ref{tab:success_rates} aggregates these outcomes into success rates, and the gap between its two columns is the central reliability finding. The proprietary models complete the entire workflow on most runs, GPT-5 on 80 percent and o4-mini-high on 57 percent, and almost always produce a functionally correct baseline, at 97 and 93 percent. The open-weights models are far less likely to finish the optimization loop, gpt-oss-120b on 37 percent and Qwen3.5-397B-A17B on 27 percent, yet both still produce a correct baseline on roughly two-thirds of runs. The open-weights models are therefore frequently capable of a correct translation but fragile through the iterative optimization that follows, as the per-kernel trajectories below make concrete.

\begin{figure}[htbp]
    \centering
    \includegraphics[width=\linewidth,trim= 0 0 0 10,clip]{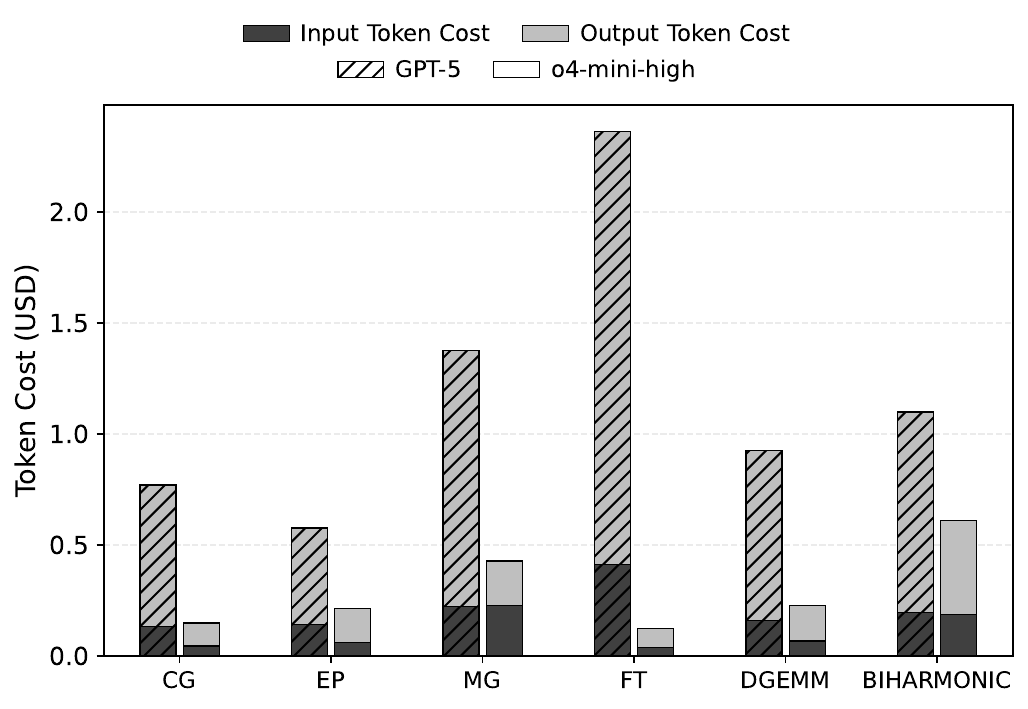}
    \caption{Mean OpenAI API token cost per kernel on the NVIDIA A100 for GPT-5 and o4-mini-high, averaged over the five runs. Each bar is stacked into input token cost (dark) and output token cost (light) and covers the full workflow, including the baseline and all optimization rounds.}
    \label{fig:token_costs}
\end{figure}

Fig.~\ref{fig:token_costs} reports the OpenAI API token cost for the two proprietary models, split into input and output cost per kernel. The open-weights models are self-hosted and incur no per-token cost, so they are not plotted here. GPT-5 is consistently more expensive than o4-mini-high on every kernel, and the gap is driven mainly by output tokens, the lighter portion of each GPT-5 bar, which reflects the volume of code it emits across the optimization rounds. FT is the most expensive kernel for GPT-5 at just over two dollars, while o4-mini-high stays well under one dollar throughout. Even so, both models complete the full translation and optimization of every kernel for only a few U.S. dollars, keeping the autonomous approach economically practical.

\begin{figure*}[htbp]
    \centering
    \includegraphics[width=\linewidth]{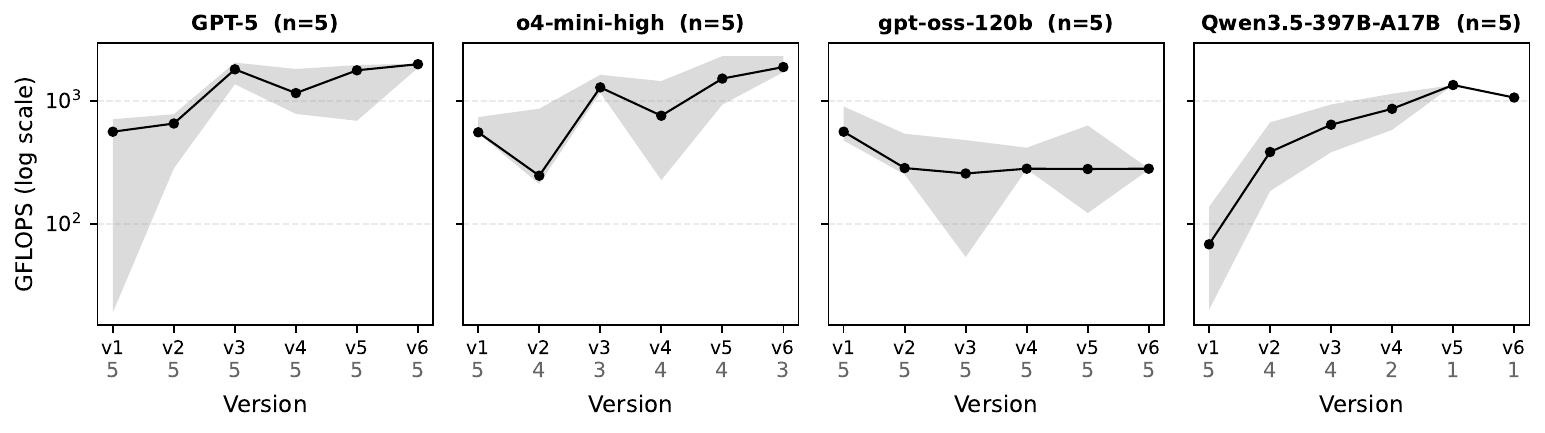}
    \caption{Optimization trajectory of the DGEMM kernel on the NVIDIA A100, measured in GFLOPS at the maximum input size across successive versions (v1 baseline through v6). Each panel corresponds to one LLM; the solid line is the median across runs and the shaded band the interquartile range. The gray number beneath each version is the count of runs (out of five) that produced a functionally correct version at that stage.}
    \label{fig:optimization_dgemm}
\end{figure*}

Fig.~\ref{fig:optimization_dgemm} shows the optimization trajectory for DGEMM, measured in GFLOPS at the maximum input size across successive versions. DGEMM is a clean basis for comparing how the optimization loop behaves across models, since all four produce a baseline on every run. GPT-5 improves steadily and with a narrow interquartile band, all five runs survive to v6, giving it the most consistent optimization behavior of any model. o4-mini-high reaches a comparable peak by v6 but along a non-monotonic path with a much wider band, and its fastest version is not always the last, a direct illustration of the non-deterministic optimization that motivates retaining every version rather than only the final one. gpt-oss-120b completes all five runs but does not improve on its baseline; its work-rate drops below the v1 value after the first optimization round and then stays flat, so the rounds yield no gain. Qwen3.5-397B-A17B starts from a much lower baseline and climbs over successive versions, but its surviving runs fall to one by v6 and it finishes below both proprietary models.

\begin{figure*}[htbp]
    \centering
    \includegraphics[width=\linewidth]{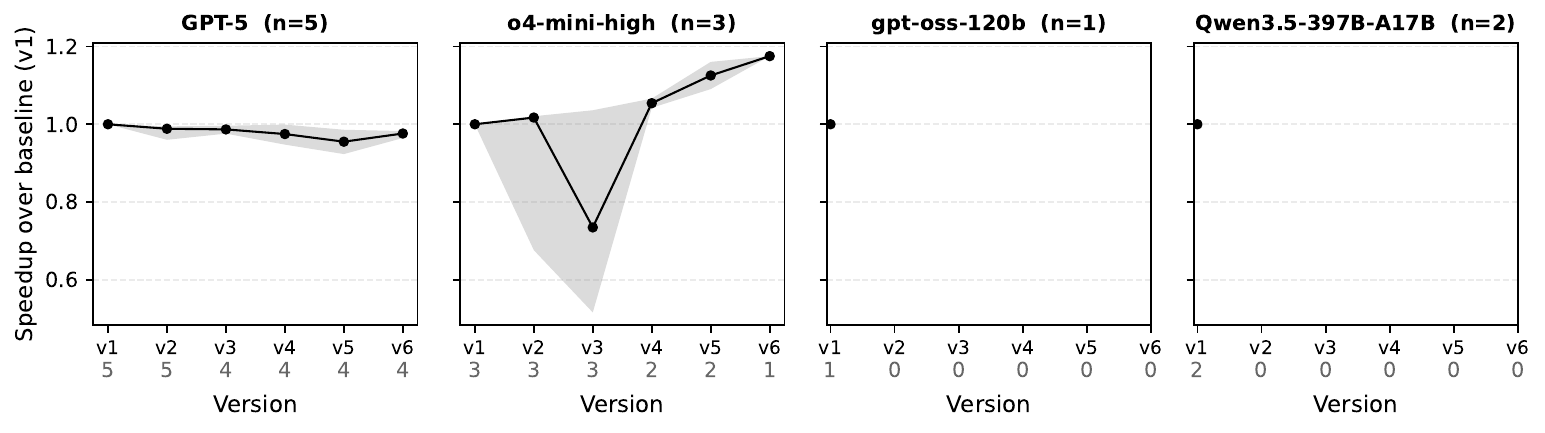}
    \caption{Optimization trajectory of the biharmonic kernel on the NVIDIA A100, reported as wall-clock speedup over the baseline (v1) at the maximum input size. The panel layout, median line, interquartile band, and per-version surviving run counts follow Fig.~\ref{fig:optimization_dgemm}. The n in each panel title is the number of runs, out of five, that produced a functionally correct baseline.}
    \label{fig:optimization_biharmonic}
\end{figure*}

Fig.~\ref{fig:optimization_biharmonic} shows the same view for the biharmonic kernel, reported as wall-clock speedup over the baseline since the kernel has no canonical FLOP count to anchor a GFLOPS figure (Section~\ref{sec:experimentalsetup}). The contrast with DGEMM is sharp, and it begins with the baselines: GPT-5 produces a correct baseline on all five runs, o4-mini-high on three, gpt-oss-120b on one, and Qwen3.5-397B-A17B on two, so this real-world multi-file kernel is the hardest for the workflow to even start. GPT-5 completes the optimization rounds on most runs but stays at or just below its baseline, so optimization yields little and the fastest retained version is frequently the baseline itself. o4-mini-high regresses by about a quarter at v3 before recovering to roughly a 1.17x speedup by v6, but only a single run survives that far, so the final gain rests on one trajectory. The open-weights models rarely produce even a baseline, and on the runs that do, they fail every subsequent optimization round. Together with DGEMM, these trajectories show that optimization gains are kernel-dependent rather than guaranteed, and that the open-weights models are the first to fall away as the kernel grows more demanding.

\begin{figure}[htbp]
    \centering
    \includegraphics[width=\linewidth,trim= 0 0 0 10,clip]{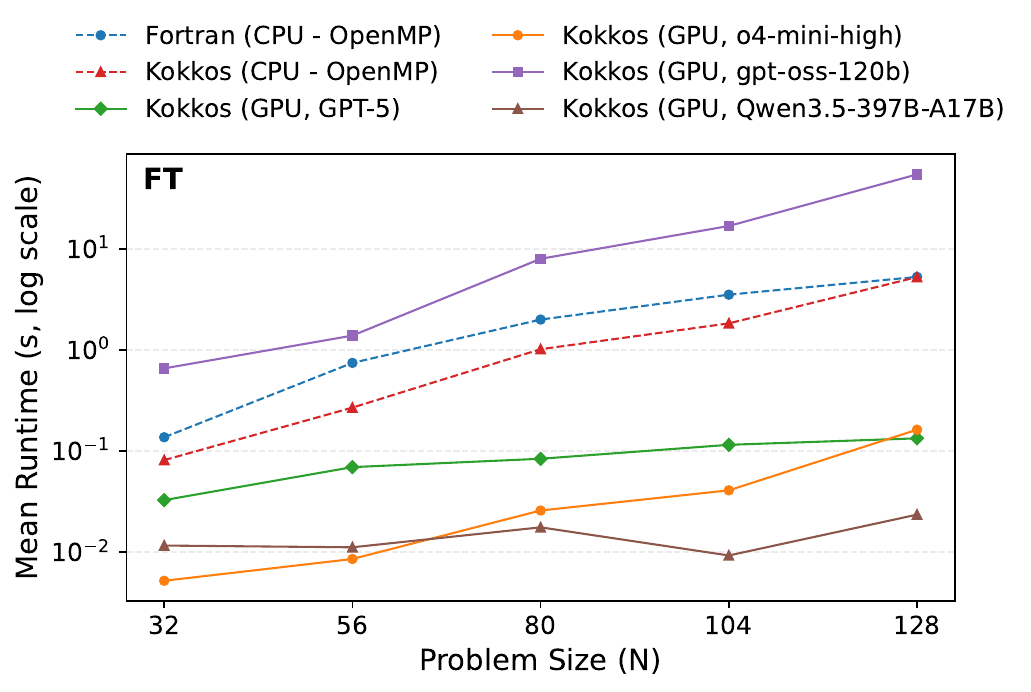}
    \caption{Mean runtime of the FT kernel on the NVIDIA A100 against problem size, with runtime shown on a logarithmic scale, comparing the Fortran CPU (OpenMP) baseline, the Kokkos CPU (OpenMP) translation from the best version of GPT-5, and the best Kokkos GPU version produced by different LLMs. Each line is the mean over the five runs.}
    \label{fig:runtime_comparison_ft_a100}
\end{figure}

Fig.~\ref{fig:runtime_comparison_ft_a100} compares FT runtimes on the A100 across the Fortran CPU baseline, the Kokkos CPU translation from GPT-5's best version, and the best Kokkos GPU version from each model. Two points stand out. First, the Kokkos CPU translation is already faster than the original Fortran CPU version at every problem size, so a benefit accrues from modernization alone, before any GPU is involved. Second, the GPU versions separate the models clearly: GPT-5, o4-mini-high, and Qwen3.5-397B-A17B all run far faster than either CPU version, whereas gpt-oss-120b produces the slowest GPU code, which at the larger problem sizes is slower than even the Fortran CPU baseline. This is a case where an open-weights model produces functionally correct but poorly performing code, a distinction the success rate alone would not capture. Representative optimizer diffs and categorized open-weights failure traces (compilation errors, invalid-semantics hallucinations, and threshold exhaustion) are provided in the supplementary material.

Finally, Fig.~\ref{fig:runtime_comparison_cg_mi250} extends the evaluation to the AMD MI250 using the CG kernel. Run on the MI250 partition, the workflow autonomously produces Kokkos code that compiles and runs on the AMD GPU through the HIP backend. All three models from the single-run experiments generate GPU code that outperforms the Fortran CPU baseline across the full range of problem sizes, with o4-mini-high fastest, GPT-5 next, and Llama 4 Maverick slowest. The same workflow also completes end-to-end on the ARM-based GH200, with the full per-architecture results provided in the supplementary material. Together with the A100 matrix, this shows that the workflow generates functional and performant Kokkos code across NVIDIA and AMD GPUs and an ARM host, with quantitative performance characterized on the A100 and MI250.

\begin{table*}[htbp]
\renewcommand{\arraystretch}{1.15}
\centering
\scriptsize
\begin{threeparttable}
\caption{Roofline analysis (double precision) for the best-performing version of each kernel on the NVIDIA A100.}
\label{tab:roofline_analysis}
\begin{tabularx}{\textwidth}{|L{1.2}|C{1.2}|C{1.0}|C{1.0}|C{1.0}|C{0.6}|}
\hline
\makecell[l]{\textbf{Kernel (s = input size)}} & \textbf{Best LLM} &
\makecell{\textbf{Achieved Performance}} & \makecell{\textbf{Arithmetic Intensity} \\ \textbf{(FLOP/byte)}} &
\makecell{\textbf{\% of Roof Achieved}} & \textbf{Bound} \\
\hline
CG ($s = 10^6$)         & Qwen3.5-397B-A17B & 998.89 GFLOPS & 0.839   & 76.54\% & Memory  \\ \hline
EP ($\log_2 s = 28$)    & GPT-5             & 6.23 TFLOPS   & $>10^4$ & 83.07\% & Compute \\ \hline
MG ($s = 256$)          & o4-mini-high      & 2.80 TFLOPS   & 2.898   & 62.11\% & Memory  \\ \hline
FT ($s = 128$)          & Qwen3.5-397B-A17B & 268.50 GFLOPS & 1.021   & 16.91\% & Memory  \\ \hline
DGEMM ($s = 4096$)      & Qwen3.5-397B-A17B & 2.63 TFLOPS   & 11.666  & 35.12\% & Compute \\ \hline
BIHARMONIC ($s = 2048$) & GPT-5             & 2.91 TFLOPS   & 16.895  & 11.07\% & Memory  \\ \hline
\end{tabularx}
\begin{tablenotes}
\raggedright
\item \scriptsize \textbf{Note:} Performance and arithmetic intensity are hardware-measured (profiler) rates, not the algorithmic FLOP model used elsewhere; for biharmonic, which has no canonical FLOP count, only this measured rate is meaningful. GPU FP64 peak $\sim$7.5 TFLOPS, HBM peak $\sim$1.555 TB/s, ridge point $\sim$4.82 FLOP/byte (1$\times$ A100 40GB SXM4). The \% of roof reports the fraction of the memory-bandwidth bound for memory-bound kernels and of peak FP64 for compute-bound kernels.
\end{tablenotes}
\end{threeparttable}
\end{table*}

The roofline analysis in Table~\ref{tab:roofline_analysis} reports the best-performing version of each kernel across the four models. The compute-bound kernels reach the highest utilization, with EP at 83.07 percent of peak FP64 and DGEMM at 35.12 percent. Among the memory-bound kernels, CG and MG reach 76.54 and 62.11 percent of the memory-bandwidth roof, while FT and biharmonic are much lower at 16.91 and 11.07 percent. For FT, the gap reflects the difficulty of automatically restructuring data movement, exploiting the cache hierarchy, and acting on profiler feedback aimed at memory behavior; for biharmonic, the measured time is dominated by per-repetition allocation and initialization rather than the operator itself. The best version is not always from a proprietary model: Qwen3.5-397B-A17B yields the top CG, FT, and DGEMM versions, so although the open-weights models complete far fewer runs, an individual successful version can match or exceed the proprietary models, locating the gap in reliability rather than peak attainable performance. Sustaining high fractions of peak bandwidth or compute in Kokkos is difficult even for experienced programmers, making the levels reached here by an autonomous pipeline noteworthy.

Taken together, these results establish that an agentic AI workflow can autonomously translate, optimize, and deploy legacy Fortran kernels as portable Kokkos implementations across NVIDIA and AMD GPUs and an ARM host. The proprietary models are the most reliable, completing the full workflow on the majority of runs, with GPT-5 and o4-mini-high producing comparable results that trade places depending on the kernel. The open-weights models complete far fewer runs and collapse on specific Kokkos abstractions, yet when they do succeed their best versions are competitive and in several cases the fastest measured, so the current gap is primarily one of reliability rather than peak capability.


\section{Threats to Validity}
\label{sec:threats}

We discuss the main threats to validity and the steps taken to control them. For construct validity, functional correctness is established by the Functionality Tester, which compares each version against the Fortran reference within a numerical tolerance (Section~\ref{sec:methods}). Two kernels need special handling: EP is verified through its statistical output, and the biharmonic reference uses a deterministic, index-based initialization so the Fortran and Kokkos sides start from identical inputs, a change confined to input setup. A separate concern is reward hacking, where an optimized version satisfies the tolerance while computing less than the full kernel; we observed this in a few Qwen3.5-397B-A17B DGEMM versions whose reported rate exceeded the hardware peak, and we exclude any such physically impossible result so the reported rates reflect genuine computation. Since the oracle uses fixed inputs, randomizing them across trials is a natural next step that would let the same check detect this directly.

For internal and conclusion validity, because LLM outputs are stochastic, the A100 evaluation uses five runs per model and kernel and reports medians and interquartile ranges; the MI250 and GH200 evaluations are single-run and serve as portability evidence, not statistical comparisons. The repair thresholds, open-weights sampling settings, and profiler feedback were held fixed across configurations to keep the comparison even, and studying their individual effect is left to future work.

For external validity, the six kernels span compute- and memory-bound regimes and single- and multi-file structure, but do not capture larger production applications with distributed-memory parallelism, global state, I/O, build-system integration, and richer control flow, which is a clear next step, and the hardware and model set are a snapshot of a fast-moving area. We therefore expect the qualitative conclusions, in particular the reliability gap between proprietary and open-weights models, to hold, while the absolute figures are best read as a characterization of present-day models rather than a fixed ranking.

For reproducibility, the archived prompts, sources, generated programs, and per-run data can be inspected and recompiled exactly, while new model runs reproduce only as distributions: the open-weights models have fixed, public weights but sampled outputs, and the proprietary models are served through evolving APIs whose behavior may shift, so we pin model versions (Table~\ref{tab:llms}) and report distributions over five runs. Separately, the pipeline's orchestration code cannot be released under U.S. Department of Energy export-control restrictions; the supplementary material contains all approved-for-release artifacts, and Section~\ref{sec:methods} describes the workflow in sufficient detail for re-implementation.


\section{Conclusion and Future Work}
\label{sec:conclusion}

This work demonstrates that agentic AI can autonomously modernize legacy Fortran kernels into portable, performant Kokkos C\texttt{++} across NVIDIA and AMD GPUs and an ARM host, with several kernels reaching substantial fractions of peak performance and memory bandwidth and completing each kernel in hours rather than the far longer manual porting effort. The proprietary models GPT-5 and o4-mini-high completed the full pipeline for only a few U.S. dollars, generating code that consistently outperformed the original Fortran baselines. The open-weights models completed fewer runs and more often failed on specific Kokkos abstractions, yet their best successful versions were competitive and in several cases the fastest measured, indicating that the current gap is one of reliability rather than peak capability. These results show feasibility at the kernel level rather than deployment-level reliability, since even GPT-5 completed the full workflow on only 80 percent of runs; improving this reliability is the main step toward using the pipeline without developer oversight.

Several avenues remain open for future work. First, our findings apply at the kernel level, and the kernels may overlap with public code seen during training; extending them to complete applications, with interconnected modules, global state, I/O, and build systems, will require codes that have no C\texttt{++} versions, which are often proprietary or export-controlled and harder to use reproducibly. Second, the functionality testing was similarly built as a proof-of-concept tailored to these kernels; a more general and dynamic framework is needed for larger applications, and future work could use AI agents to automatically generate and refine domain-specific tests, while randomizing test inputs across trials would let the check detect under-computation directly. Third, the optimization pipeline is effective and uses profiler feedback, but follows a sequential strategy in which each round builds on the previous one, and exploring alternatives, such as retaining only the best-performing versions at each stage, could improve consistency at the cost of additional runtime. Finally, we used the same LLM across all agents to keep the benchmarking consistent; future versions could assign specialized models per task, for example code-specialized models for translation and fixing, lightweight models for validation, and high-reasoning models for optimization.

\section*{Supplementary materials}

\label{sec:supplementary}

All artifacts are archived on Zenodo (\url{https://zenodo.org/records/22679376}). The agent prompts are included for both experiments. For the main multi-run A100 experiments, we provide the Fortran and generated Kokkos C\texttt{++} sources, per-run data, failures, performance and roofline results, run summaries, and all plots. For the supplementary single-run MI250 and GH200 experiments, we provide the Fortran and generated Kokkos C\texttt{++} sources, agent invocations, runtime settings, and plots.


\bibliographystyle{ieeetr}
\bibliography{references}


\end{document}